\documentclass[twocolumn,secnumarabic,assume,nobibnotes,aps,pre]{revtex4-2}
\usepackage{amsmath,amssymb,amsfonts}
\usepackage{graphicx}
\usepackage{xcolor}
\usepackage{subfig}
\usepackage{float}
\usepackage{siunitx}
\usepackage{parskip}
\usepackage{rotating}
\usepackage{multirow}
\usepackage{colortbl}
\usepackage[T1]{fontenc}
\usepackage{lmodern}
\usepackage{placeins}
\usepackage{booktabs}
\allowdisplaybreaks

\begin{document}

\title{Phase-controlled elastic, inelastic, and coalescent collisions of two-dimensional flat-top solitons}

\author{M. O. D. Alotaibi$^{1,*}$, Y. O. A. Abughnheim$^{2}$, L. Al Sakkaf$^{3}$, and U. Al Khawaja$^{2,4}$}
\address{$^{1}$Department of Physics, College of Science, Kuwait University, Sabah Al Salem University City, P.O. Box 5969, Safat 13060, Shadadiya, Kuwait,\\
$^{2}$Department of Physics, School of Science, The University of Jordan, Amman 11942, Jordan,\\
$^{3}$Department of Applied Sciences, College of Engineering, Abu Dhabi University, Al-Ain 59911, United Arab Emirates,\\
$^{4}$Department of Physics, United Arab Emirates University,P.O. Box 15551, Al-Ain, United Arab Emirates.}
\email{Contact author: majed.alotaibi@ku.edu.kw}

\begin{abstract}
We investigate elastic, inelastic, and coalescent collisions between
two-dimensional flat-top solitons supported by the cubic--quintic nonlinear
Schr\"odinger equation. Numerical simulations reveal distinct collision regimes
ranging from nearly elastic scattering to strongly inelastic interactions
leading to long-lived merged states. We demonstrate that the transition between
these regimes is primarily controlled by the relative phase of the solitons at
the collision point, with out-of-phase collisions suppressing overlap and
in-phase collisions promoting strong interaction. Kinetic-energy diagnostics
are introduced to quantitatively characterize collision outcomes and to identify
phase- and separation-dependent windows of elasticity. To interpret the observed
dynamics, we extract effective phase-dependent interaction potentials from
collision trajectories, providing a mechanical picture of attraction and
repulsion between flat-top solitons. The stability of merged states formed after
strongly inelastic collisions is explained by their lower energetic cost,
arising from interfacial energetics, where a balance between internal pressure
and edge tension plays a central role. A variational analysis based on direct
energy minimization supports this picture by revealing robust energetic minima
associated with stationary two-dimensional flat-top solitons.
\end{abstract}

\maketitle

\section{Introduction}
\label{sec:intro}

Solitons are localized nonlinear excitations arising from a balance between
dispersion and nonlinearity and are distinguished by their robustness under
propagation and interaction
\cite{books1,books2,books3,books4,books5,books6,books7,books8}.
In integrable one-dimensional models, such as the cubic nonlinear
Schr\"odinger equation (NLSE), solitons undergo perfectly elastic collisions:
after interaction, each soliton recovers its original shape, velocity, and
amplitude, up to a phase shift
\cite{Prinari2023,AblowitzSegur1981}.
When integrability is broken, either through competing nonlinearities or by increasing the dimensionality, soliton interactions may become inelastic, allowing for radiation emission, excitation of internal modes, deformation, or fusion\cite{abd}.

Flat-top solitons (FTSs) represent a particularly rich class of nonintegrable
localized states. They arise, for instance, in systems with competing self-focusing and
self-defocusing nonlinearities, most commonly cubic and quintic terms, and are
characterized by an extended, nearly uniform bulk density bounded by relatively
sharp edges
\cite{cheiney2018,bottcher2021,luo2021,cappellaro2017,ferioli2019,cikojevic2019,sachdeva2020,ota2020,hu2020,guo2021,guebli2021}

This structure leads to liquid-like properties in FTSs, including an effectively incompressible interior and a finite `surface' tension associated with their interfaces. Such states have been predicted and observed in a variety of
physical settings, including nonlinear optical media and ultracold atomic
systems \cite{Novoa2009}.
The collision dynamics of one-dimensional FTSs has been studied in
detail \cite{Li2018,Khawaja2024}. In particular, earlier work demonstrated that the transition between
elastic and inelastic collisions is primarily controlled by the
\emph{relative phase at the collision point}
\cite{Khawaja2024}.
Out-of-phase collisions significantly suppress overlap and lead to nearly elastic scattering, whereas in-phase collisions promote strong overlap, radiation emission, and, in
some cases, permanent merger. These regimes were quantitatively characterized
using kinetic-energy diagnostics and further interpreted in terms of effective
interaction forces, establishing a coherent picture in which interference,
nonintegrability, and energetics jointly govern the collision outcome.

Despite this progress, comparatively little is known about collisions of
\emph{two-dimensional} FTSs. Extending collision studies to higher
dimensions is not a trivial generalization. In two dimensions, flat-top
solitons possess extended interfaces whose curvature and energetic cost play a
central role in their dynamics, and additional channels for transverse
deformation and radiation become available. As a result, inelastic collisions
may lead not only to transient distortion but also to long-lived merged states,
whose stability cannot be understood solely on the basis of phase interference.

In this work we investigate collisions of two-dimensional FTSs
within the cubic--quintic nonlinear Schr\"odinger equation. Using numerical simulations, we demonstrate that, as in one dimension, the relative
phase at the collision point remains the dominant control parameter determining
whether collisions are elastic or inelastic. However, the two-dimensional
geometry enriches the inelastic dynamics through transverse deformation,
enhanced radiation, and the emergence of stable coalesced states. To
quantitatively distinguish collision regimes, we introduce kinetic-energy
diagnostics that reveal clear separation- and phase-dependent windows of
elasticity and inelasticity.

Beyond the dynamical characterization, we develop two complementary
interpretations of the observed behavior. First, we extract effective
phase-dependent interaction forces and potentials directly from collision
trajectories, providing a mechanical picture of attraction for in-phase
solitons and repulsion for out-of-phase solitons. Second, we connect the outcome
of strongly inelastic collisions to the energetic structure of stationary
two-dimensional FTSs. By combining interfacial energetics and a Young--Laplace--type balance
\cite{Novoa2009}, in which the bulk pressure pushing outward is balanced by edge
tension pulling inward, with a variational description based on direct energy
minimization, we show that merged states correspond to stable energetic minima
and are not merely transient outcomes of the collision. Together, these results establish a unified framework for understanding elastic,
inelastic, and coalescent collisions of two-dimensional FTSs,
linking phase interference, effective forces, and energetic stability in a higher-dimensional setting.

This article proceeds as follows. In Sec.~\ref{sec:model}, we introduce the
two-dimensional cubic--quintic nonlinear Schr\"odinger equation and summarize
its conserved quantities. Section~\ref{sec:stationary_init} describes the
numerical construction of stationary two-dimensional FTSs and the
initialization of two-soliton configurations via imaginary-time propagation (ITP).
Direct real-time simulations of FTS collisions are presented in
Sec.~\ref{sec:collisions}, with emphasis on the role of the relative phase at
the collision point. In Sec.~\ref{sec:KE_diagnostics}, kinetic-energy
diagnostics are introduced to quantitatively distinguish elastic, nearly
elastic, and inelastic collision regimes, and their dependence on initial
separation and phase is analyzed. Effective interaction forces and potentials
extracted from collision dynamics are discussed in
Sec.~\ref{subsec:effective_force}, providing a mechanical interpretation of
phase-controlled attraction and repulsion. The formation and stability of
long-lived merged states following strongly inelastic collisions are examined
in Sec.~\ref{sec:discussion_coalescence}, where interfacial energetics and a
Young--Laplace--type balance are used to explain their persistence. The paper
concludes in Sec.~\ref{sec:conclusions} with a summary of the main results and
an outlook for future work.

\section{Theoretical model}
\label{sec:model}

We study the two-dimensional cubic--quintic nonlinear Schr\"odinger equation
\begin{equation}
i\,\psi_{t}
+ g_{1}\left(\psi_{xx} + \psi_{yy}\right)
+ g_{2}\,|\psi|^{2}\psi
+ g_{3}\,|\psi|^{4}\psi
= 0,
\label{eq:CQNLSE2D}
\end{equation}
where $\psi(x,y,t)$ denotes the complex field envelope.
The coefficient $g_{1}$ controls the dispersive term, while $g_{2}>0$ and $g_{3}<0$ represent cubic self-focusing and quintic self-defocusing nonlinearities, respectively. The balance between dispersion and nonlinear response enables localization, while the competition between cubic self-focusing and quintic self-defocusing nonlinearities gives rise to flat-top soliton profiles.

Equation~(\ref{eq:CQNLSE2D}) is nonintegrable, and as a consequence soliton interactions are not constrained to be perfectly elastic. Collisions may therefore involve radiation emission, waveform deformation, or permanent merger, depending on the interaction parameters. This intrinsic nonintegrability underlies the inelastic collision dynamics examined in the following sections. Stationary FTSs are sought in the form
\begin{equation}
\psi(x,y,t)=u(x,y)\,e^{-i\mu t},
\end{equation}
which leads to the nonlinear elliptic equation
\begin{equation}
\mu u
+ g_{1} \left(u_{xx}+u_{yy}\right)
+ g_{2} u^{3}
+ g_{3} u^{5}
= 0 .
\label{eq:stationary_CQ}
\end{equation}
In the flat-top regime, the solution $u(x,y)$ exhibits an approximately constant density in the bulk, with the nonlinear terms nearly balancing,
\begin{equation}
g_{2} u^{2} + g_{3} u^{4} \approx -\mu,
\end{equation}
while gradient contributions become significant only near the soliton boundary.
Because Eq.~\eqref{eq:stationary_CQ} does not admit closed-form localized solutions in two dimensions, stationary states are obtained numerically, as described in Sec.~\ref{sec:stationary_init}. The conserved quantities associated with Eq.~\eqref{eq:CQNLSE2D} are the norm,
\begin{equation}
N = \int_{-\infty}^{\infty}\!\!\int_{-\infty}^{\infty}
|\psi|^{2}\,dx\,dy,
\label{eq:norm}
\end{equation}
the momentum,
\begin{equation}
P = (P_x,P_y),
\qquad
P_{x,y} =
\int_{-\infty}^{\infty}\!\!\int_{-\infty}^{\infty}
\mathrm{Im}\!\left(\psi^{*}\psi_{x,y}\right)\,dx\,dy,
\label{eq:momentum}
\end{equation}
and the total energy,
\begin{align}
E &=
\int_{-\infty}^{\infty}\!\!\int_{-\infty}^{\infty}
\Bigg[
g_{1} \left(|\psi_x|^{2}+|\psi_y|^{2}\right)
-\frac{1}{2}g_{2}|\psi|^{4} \notag\\&
-\frac{1}{3}g_{3}|\psi|^{6}
\Bigg]
\,dx\,dy .
\label{eq:total_energy}
\end{align}

These conserved quantities provide the energetic and dynamical framework used to characterize collision outcomes in the subsequent analysis.

%
\section{Stationary Flat-top Solitons and Initialization}
\label{sec:stationary_init}

\begin{figure}[!t]
    \centering
    \includegraphics[width=\columnwidth]{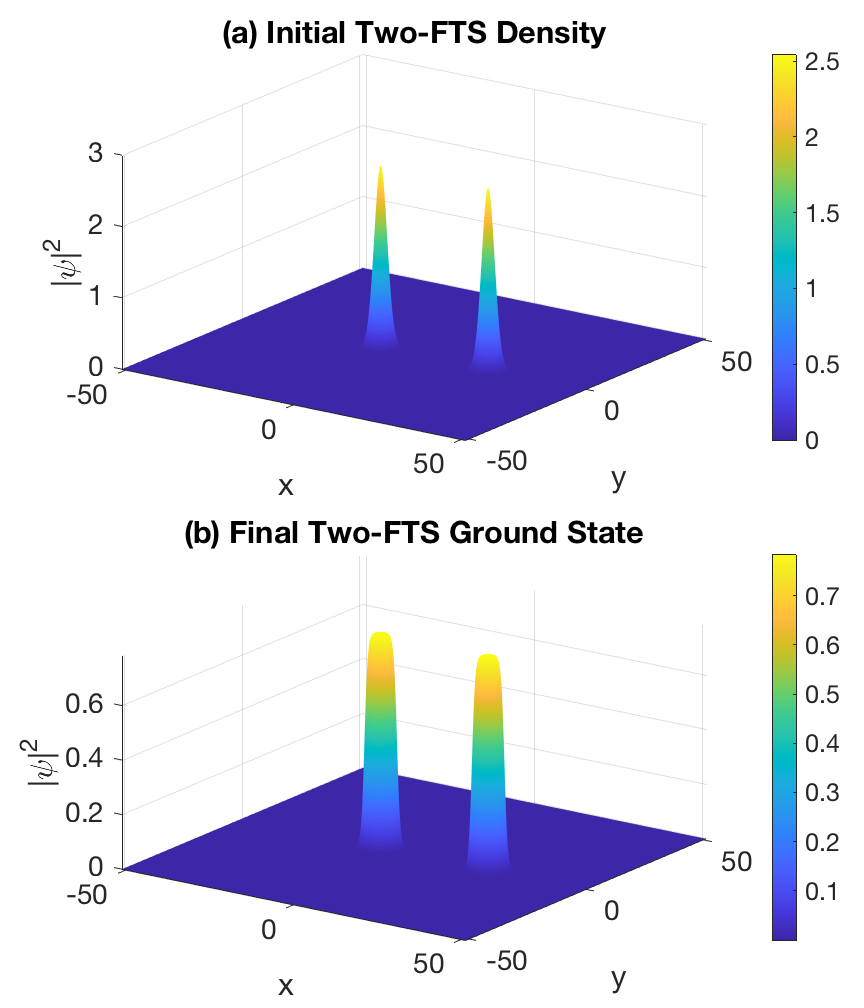}
    \caption{Initial and final density surfaces used in the numerical setup.
    (a) Initial two-Gaussian seed of width \(w = 10\), centered at
    \((x_1,y_1)=(-10,0)\) and \((x_2,y_2)=(22,0)\).
    (b) Stationary two-FTS state obtained after ITP.}
    \label{fig:twoFTS_initial_final}
\end{figure}

\begin{figure}[!t]
    \centering
    \includegraphics[width=\columnwidth]{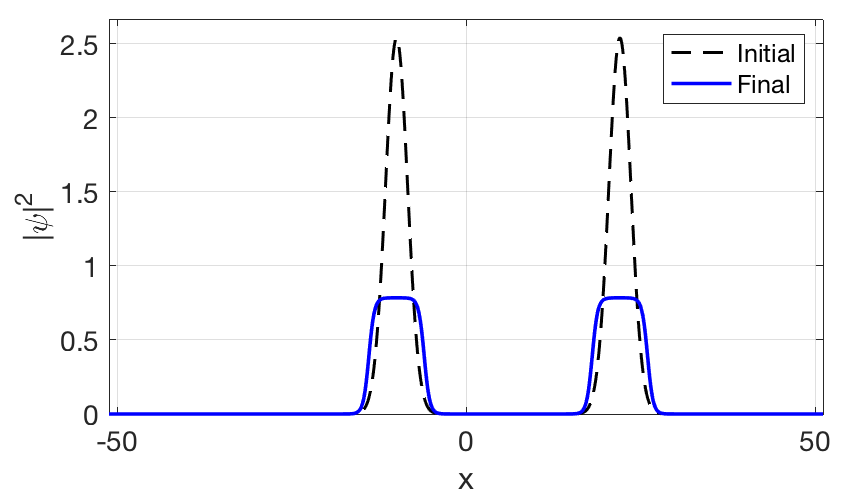}
    \caption{Central slice comparison of the initial two-Gaussian seed and the stationary
    two-FTS state produced by ITP, for the same parameters as in Fig.~\ref{fig:twoFTS_initial_final}.}
    \label{fig:centerSlice_twoFTS}
\end{figure}

Stationary two-dimensional FTSs are obtained numerically by ITP of Eq.~\eqref{eq:CQNLSE2D} using a split-step Fourier method.
Throughout this work we set $g_{1}=1/2$ and fix the nonlinear coefficients to $g_2=4$ and $g_3=-4$, and impose a target norm $N=80$.
The ITP evolution converges to the minimum-energy state at fixed norm, yielding radially symmetric FTSs. To construct a stationary two-soliton configuration, the initial field is taken as a superposition of two broad Gaussian seeds of common width $w=10$, namely
\begin{equation}
\psi(x,y,0)
=
\sum_{j=1}^{2}
\exp\!\left[
-\frac{(x-x_j)^2+(y-y_j)^2}{w^2}
\right],
\label{eq:twoGaussianSeed}
\end{equation}
where $(x_1,y_1)=(-10,0)$ and $(x_2,y_2)=(22,0)$. The field is normalized to the target norm and evolved in imaginary time until convergence. The relaxed state obtained after ITP represents the two–flat-top–soliton ground state.
Figure~\ref{fig:twoFTS_initial_final} illustrates a representative ITP relaxation from the initial two-Gaussian seed to the stationary two-FTS configuration, while Fig.~\ref{fig:centerSlice_twoFTS} shows a central density slice comparing the initial and final states.

\section{Collision dynamics}
\label{sec:collisions}

In this section we examine collisions between two-dimensional FTSs with an emphasis on the role of the \emph{relative phase} at the collision point.
All simulations are performed using a Fourier split-step method on a $1024\times1024$ grid with spatial steps $dx=dy=0.1$ and time step $dt=5\times10^{-4}$, under periodic boundary conditions. To isolate the phase-controlled interaction mechanism, we keep all numerical and physical parameters fixed and vary only the initial horizontal separation between the solitons, which determines the accumulated relative phase at impact.

Both simulations use identical velocity assignments,
with the first soliton initially at rest
($V_{1x}=V_{1y}=0$)
and the second soliton moving toward it with velocity
$V_{2x}=-0.3$ and $V_{2y}=0$,
so that the collision occurs along a straight line.
The motion of the second soliton is generated by applying an initial
phase imprinting (kick) to its stationary profile within the
two–soliton configuration obtained by ITP.
The resulting initial condition for the real-time evolution is taken as
\begin{equation}
\psi(x,y,0)
=
\sum_{j=1}^{2}
u_j(x-x_j,y-y_j)\,
\exp\!\left[i\left(V_{jx} x + V_{jy} y\right)\right],
\label{eq:initial_kick}
\end{equation}
where $u_j(x,y)$ denotes the stationary flat-top soliton, and $(V_{jx},V_{jy})$ specify the imposed
velocities of the individual solitons, as described above. The only change between the two collision scenarios is the initial location of the second soliton:
\[
(x_{2},y_{2})=
\begin{cases}
(30,0), & \text{Case (a): elastic scattering},\\[4pt]
(15,0), & \text{Case (b): inelastic scattering}.
\end{cases}
\]

\begin{figure}[t!]
    \centering
    \includegraphics[width=\columnwidth]{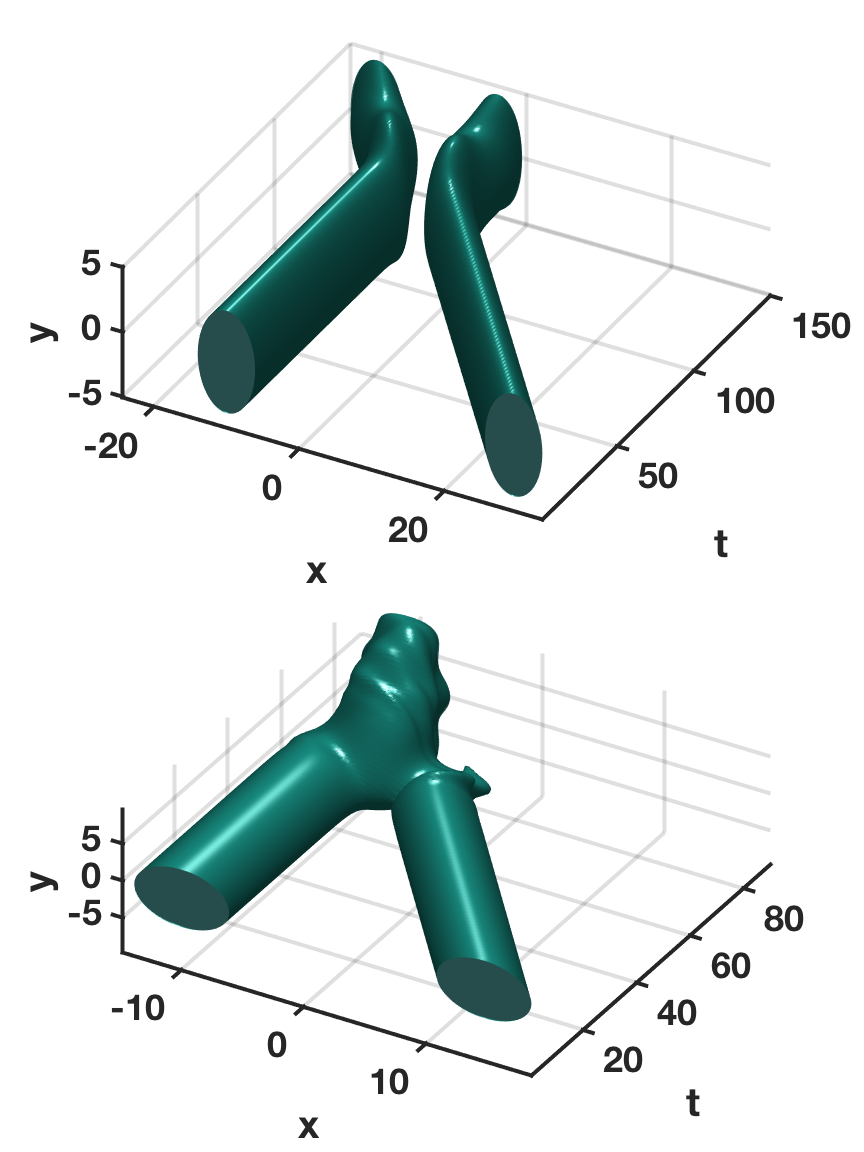}
    \caption{Comparison of two collision scenarios obtained with identical parameters, differing only in the initial $x$-position of the second soliton.
(a)~Elastic scattering for $(x_{2},y_{2})=(30,0)$.
(b)~Inelastic scattering for $(x_{2},y_{2})=(15,0)$.
The qualitative difference arises from the relative phase at the collision point.}
    \label{fig:elasticVSinelastic}
\end{figure}

\begin{figure}[t!]
    \centering
    \includegraphics[width=\columnwidth]{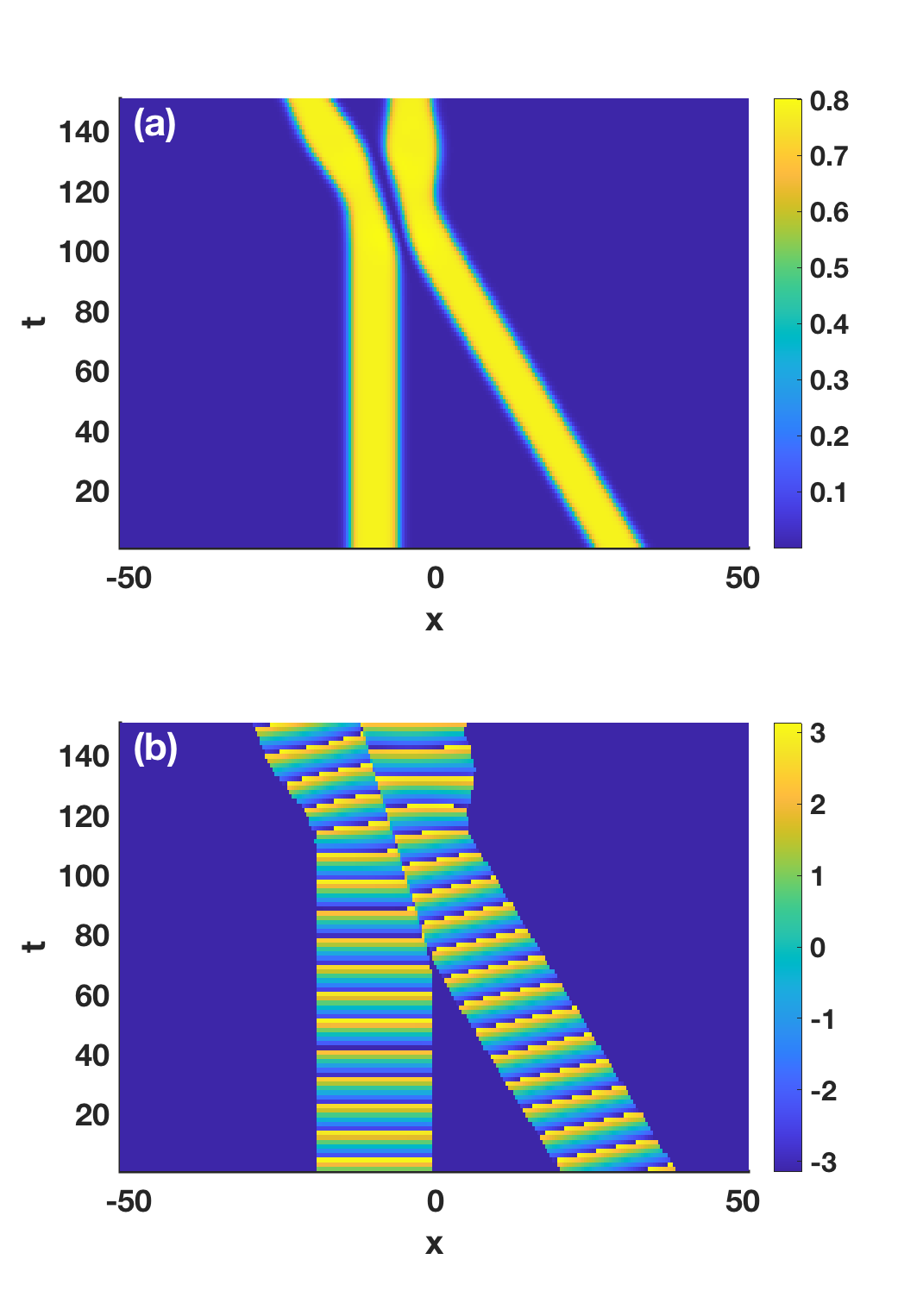}
    \caption{Density and phase evolution along $y=0$ for the elastic collision. (a) Density shows negligible overlap. (b) Phase shows a relative phase close to $\pi$ at the collision point.}
    \label{fig:elastic_phase_density}
\end{figure}

\begin{figure}[t!]
    \centering
    \includegraphics[width=\columnwidth]{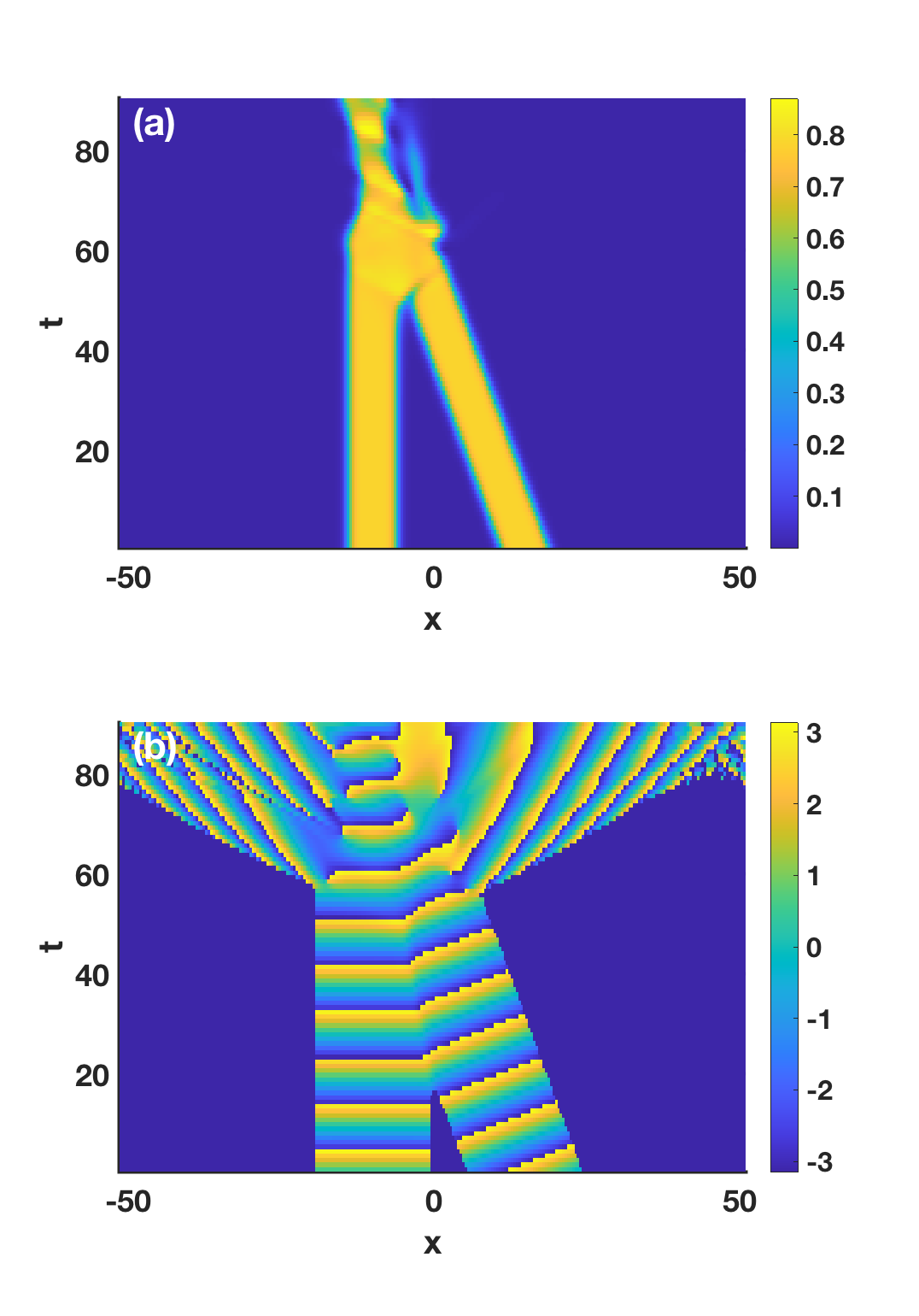}
    \caption{Density and phase evolution along $y=0$ for the \textbf{inelastic} collision. (a) Density shows strong overlap. (b) Phase shows a relative phase close to $0$ at the collision point, leading to deformation and radiation.}
    \label{fig:inelastic_phase_density}
\end{figure}

To connect the two-dimensional collision dynamics shown in Fig.~\ref{fig:elasticVSinelastic} with the underlying interaction mechanism, we extract one-dimensional slices along $y=0$ of both the density and the phase.
These slices reveal the degree of soliton overlap and the relative phase at the collision point.
Figures~\ref{fig:elastic_phase_density} and \ref{fig:inelastic_phase_density} show the corresponding density $|\psi(x,t)|^{2}$ and phase evolution.

In the \emph{elastic} case $(x_{2},y_{2}) = (30,0)$, the solitons reach the interaction region with a phase difference close to $\pi$, as evidenced by the phase evolution shown in Fig.~\ref{fig:elastic_phase_density}, which suppresses overlap and energy exchange.
In the \emph{inelastic} case $(x_{2},y_{2})=(15,0)$, the solitons are nearly in phase at impact, as shown in Fig.~\ref{fig:inelastic_phase_density}, overlap strongly, and emit radiation accompanied by long-lasting deformation.
Since all parameters are identical except for the initial separation, these results demonstrate that the \emph{relative phase at impact} is the dominant control parameter for elastic versus inelastic collisions.

%

\section{Kinetic-energy diagnostics of elastic and inelastic collisions}
\label{sec:KE_diagnostics}

To quantitatively distinguish elastic from inelastic collision outcomes, we monitor the kinetic energy of the system before and after the collision.
In an elastic collision, the kinetic energy is conserved, whereas in an inelastic collision a finite fraction of kinetic energy is irreversibly transferred to radiation and internal excitations.
Accordingly, we introduce the change in kinetic energy,
\begin{equation}
\Delta KE = KE_{\mathrm{after}} - KE_{\mathrm{before}},
\label{eq:DeltaKE_def}
\end{equation}
as a global diagnostic of collision elasticity.
Values of $\Delta KE$ close to zero indicate nearly elastic collisions, while larger deviations signal inelastic behavior.

\subsection{Separation-dependent kinetic-energy diagnostics}
\label{subsec:KE_vs_sep}

We first examine how the collision elasticity varies with the initial separation
$\Delta x = x_2 - x_1$ between the two FTSs, while keeping all other
parameters fixed.
In all cases the first soliton is held fixed at $x_1=-10$, as in the initialization
described in Sec.~\ref{sec:stationary_init}, and the separation is varied by changing
the position $x_2$ of the second soliton.
For each value of $\Delta x$, a stationary two-soliton initial state is prepared
by ITP, followed by real-time evolution with identical
collision velocities. The change in kinetic energy $\Delta KE$, defined in
Eq.~\eqref{eq:DeltaKE_def}, is used here as a quantitative diagnostic of the
collision outcome.

\begin{figure}[t]
\centering
\includegraphics[width=0.95\linewidth]{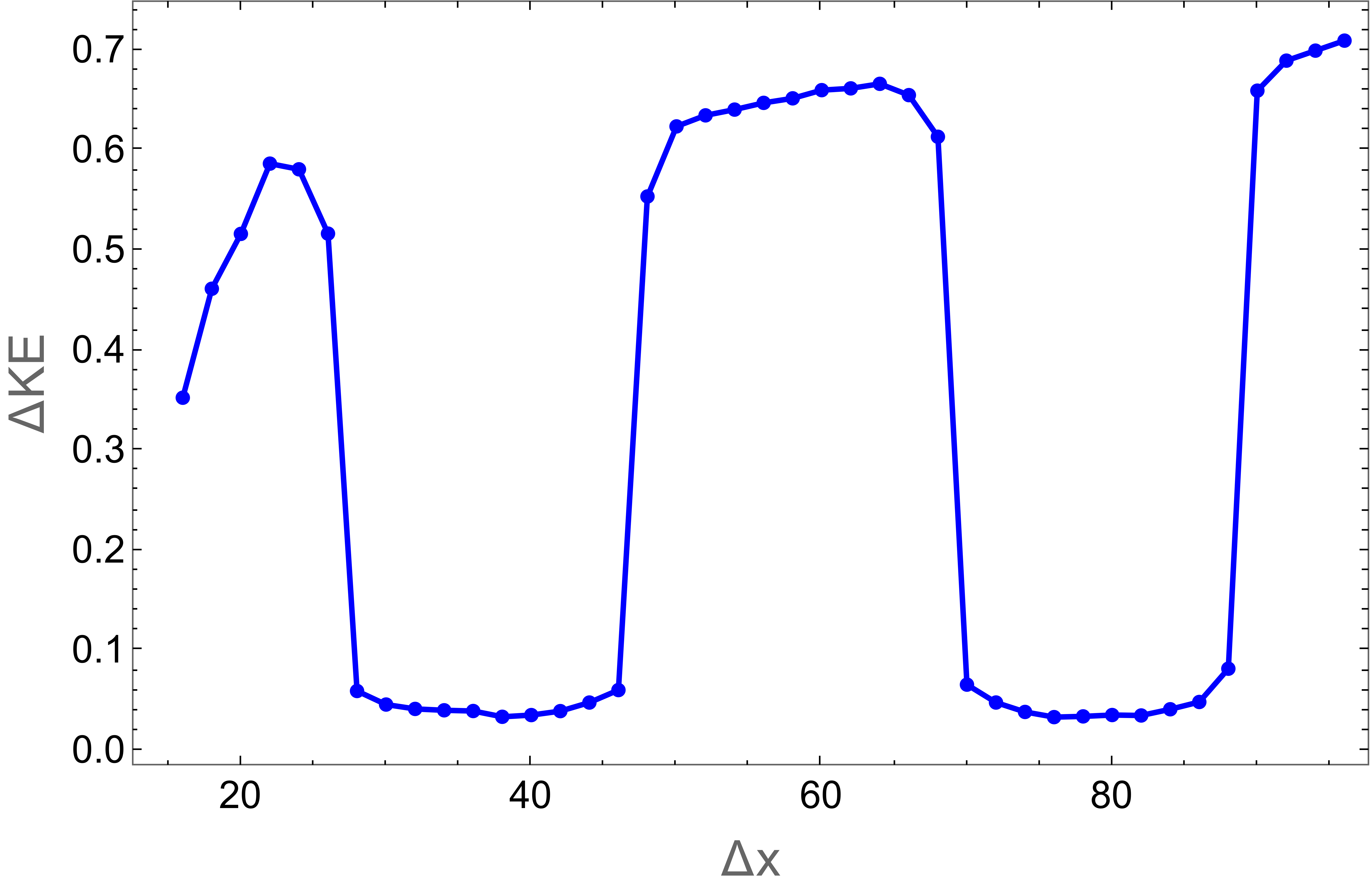}
\caption{Change in kinetic energy $\Delta KE$ as a function of the initial separation
$\Delta x = x_2 - x_1$ between two FTSs.
A clear low-$\Delta KE$ window is observed over an intermediate range of $\Delta x$,
indicating nearly elastic collisions, while larger values of $\Delta KE$ correspond
to increasingly inelastic behavior.}
\label{fig:DeltaKE_vs_sep}
\end{figure}

Figure~\ref{fig:DeltaKE_vs_sep} shows that a clear low-$\Delta KE$ window
appears over an intermediate range of $\Delta x$, indicating nearly elastic
collisions, while both smaller and larger separations lead to significantly
larger $\Delta KE$ and strongly inelastic behavior. This dependence on $\Delta x$ reflects the role of the accumulated relative phase
at the collision point.
Varying the initial separation effectively tunes the phase difference at impact,
leading to alternating elastic and inelastic collision windows.

\subsection{Phase-dependent kinetic-energy diagnostics}
\label{subsec:KE_vs_phase}

In an alternative approach, we examine the dependence of the collision elasticity
on the \emph{initially imposed} phase difference $\Delta\phi$ between the two
FTSs, while keeping all other physical and numerical parameters fixed. In particular, the soliton velocities, initial separation, norm, and nonlinear coefficients are identical for all runs in this subsection.
The parameter $\Delta\phi$ is introduced by imprinting a uniform phase shift on one of the solitons in the stationary two-soliton initial state prior to real-time evolution.

We emphasize that $\Delta\phi$ denotes the relative phase at the initial time $t=0$.
During propagation, each soliton acquires additional phase, so the phase difference at the collision point is generally different from the initially imposed value.
Nevertheless, since all other parameters are held fixed, the phase difference at impact is a deterministic function of $\Delta\phi$, allowing the collision dynamics to be parametrized in terms of the initial phase difference. For each value of $\Delta\phi$, the system is evolved in real time using the same
collision protocol, and the collision outcome is quantified by computing the
change in kinetic energy $\Delta KE$, defined in Eq.~\eqref{eq:DeltaKE_def},
before and after the collision.

\begin{figure}[t]
\centering
\includegraphics[width=0.95\linewidth]{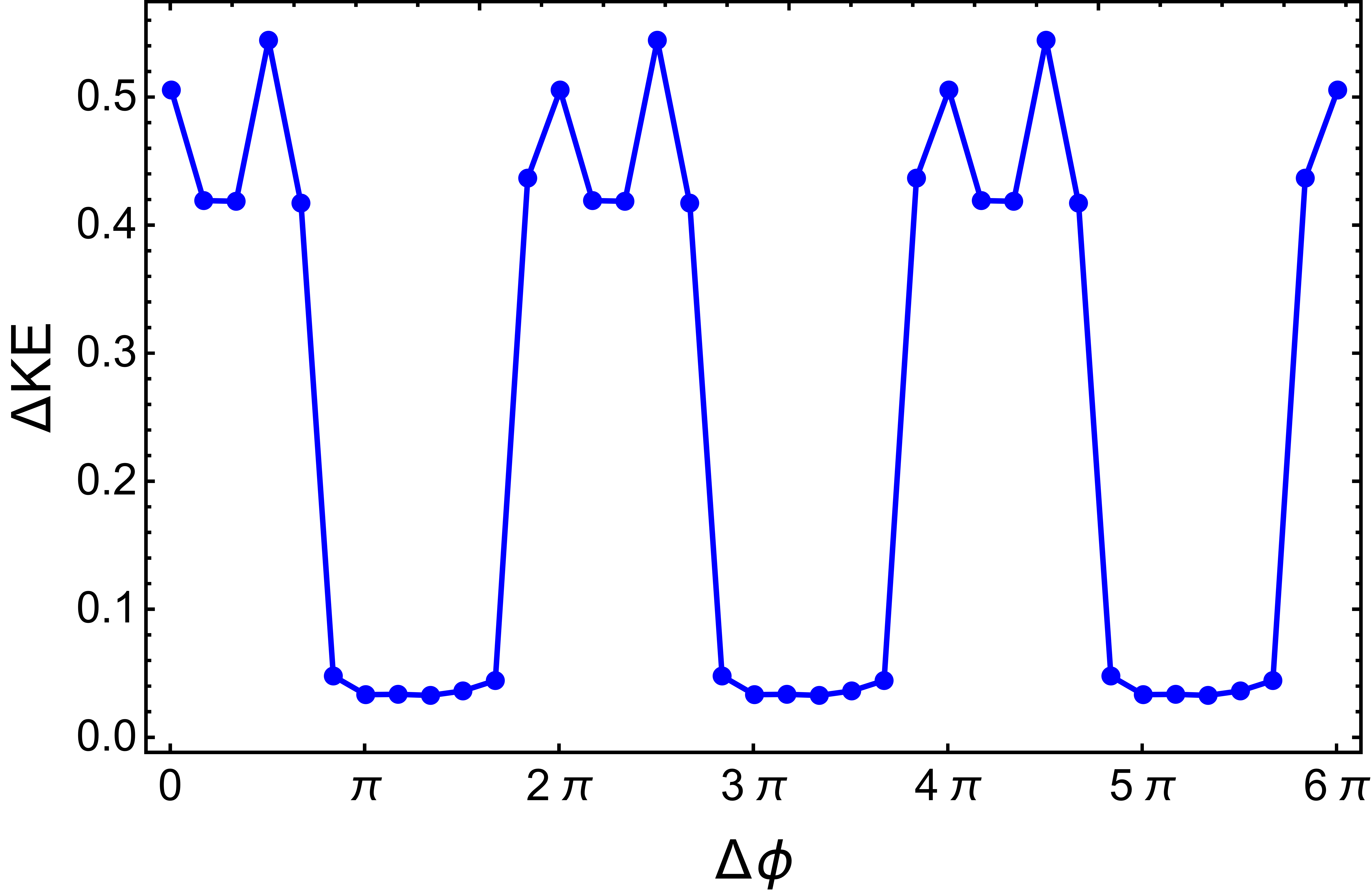}
\caption{Change in kinetic energy $\Delta KE$ as a function of the \emph{initially imposed} relative phase difference $\Delta\phi$ between two FTSs.
The parameter $\Delta\phi$ labels the phase difference at $t=0$; the relative phase at the collision point is modified during propagation by dynamical phase accumulation.
Nearly elastic collisions, characterized by $\Delta KE \approx 0$, occur for initial phases leading to out-of-phase configurations at impact, in which the two solitons remain distinct after collision.
In contrast, large values of $\Delta KE$ correspond to strongly inelastic collisions, where the solitons merge into a single flat-top structure.
The periodic dependence on $\Delta\phi$ highlights the dominant role of phase in
controlling the collision outcome.}
\label{fig:DeltaKE_vs_phase}
\end{figure}

Figure~\ref{fig:DeltaKE_vs_phase} shows the dependence of $\Delta KE$ on the initial
phase difference $\Delta\phi$. The collision outcome exhibits a periodic
dependence on $\Delta\phi$. Initial phases that lead to out-of-phase configurations at the collision point produce minimal kinetic-energy change, with the two FTSs emerging as distinct entities.
In contrast, initial phases that result in approximately in-phase collisions lead to a substantial increase in $\Delta KE$ and strongly inelastic behavior, in which the two solitons merge into a single flat-top structure.

This clear dependence on phase shows that the elasticity of collisions between two-dimensional FTSs is controlled by the relative phase at the moment of impact.
When combined with the separation-dependent results in Sec.~\ref{subsec:KE_vs_sep}, this also shows that the alternating elastic and inelastic collision windows, observed as the initial separation is varied, arise from the phase accumulated by the solitons during their propagation before collision.

In addition, we note that when all parameters are fixed and only the collision velocity is varied, the collision outcome is still determined by the relative phase at impact, as in the separation- and phase-dependent analyses.
However, increasing the velocity leads to increasingly hard collisions, and although elastic and inelastic outcomes remain clearly distinguishable, the rapid onset of strong deformation and breakup prevents the emergence of a simple periodic dependence of $\Delta KE$ on velocity.

%

\subsection{Effective interaction forces obtained from collision dynamics}
\label{subsec:effective_force}

The collision dynamics presented above demonstrate that the relative phase at
the collision point controls whether two FTSs repel, overlap, or
merge. To provide a complementary mechanical interpretation of this behavior,
we extract an effective interaction force between the solitons directly from
their real-time collision dynamics.

To this end, we consider an initial configuration in which two flat-top
solitons are placed side by side at rest. We focus on whether the interaction is
effectively attractive or repulsive as a function of the relative phase.
In particular, we track the time-dependent separation $s(t)$ between the soliton
centers of mass. The center-of-mass position of each soliton is defined through
the density-weighted coordinate
\begin{equation}
\mathbf{R}_j(t)=
\frac{\int \mathbf{r}\,|\psi_j(\mathbf{r},t)|^2\,d^2r}
{\int |\psi_j(\mathbf{r},t)|^2\,d^2r},
\qquad j=1,2,
\end{equation}
so that the instantaneous separation is given by
$
s(t)=|\mathbf{R}_2(t)-\mathbf{R}_1(t)|.
$
From the measured trajectory, an effective equation of motion may be written in
the form
\begin{equation}
M\,\ddot{s}=F(s),
\end{equation}
where $M=m/2$ is an effective reduced mass. The corresponding effective
interaction potential is obtained by integration,
\begin{equation}
V(s)=-\int_s^{\infty}F(s')\,ds',
\end{equation}
with the normalization $V(\infty)=0$ imposed numerically.

%

\begin{figure}[t]
\centering
\includegraphics[width=\columnwidth]{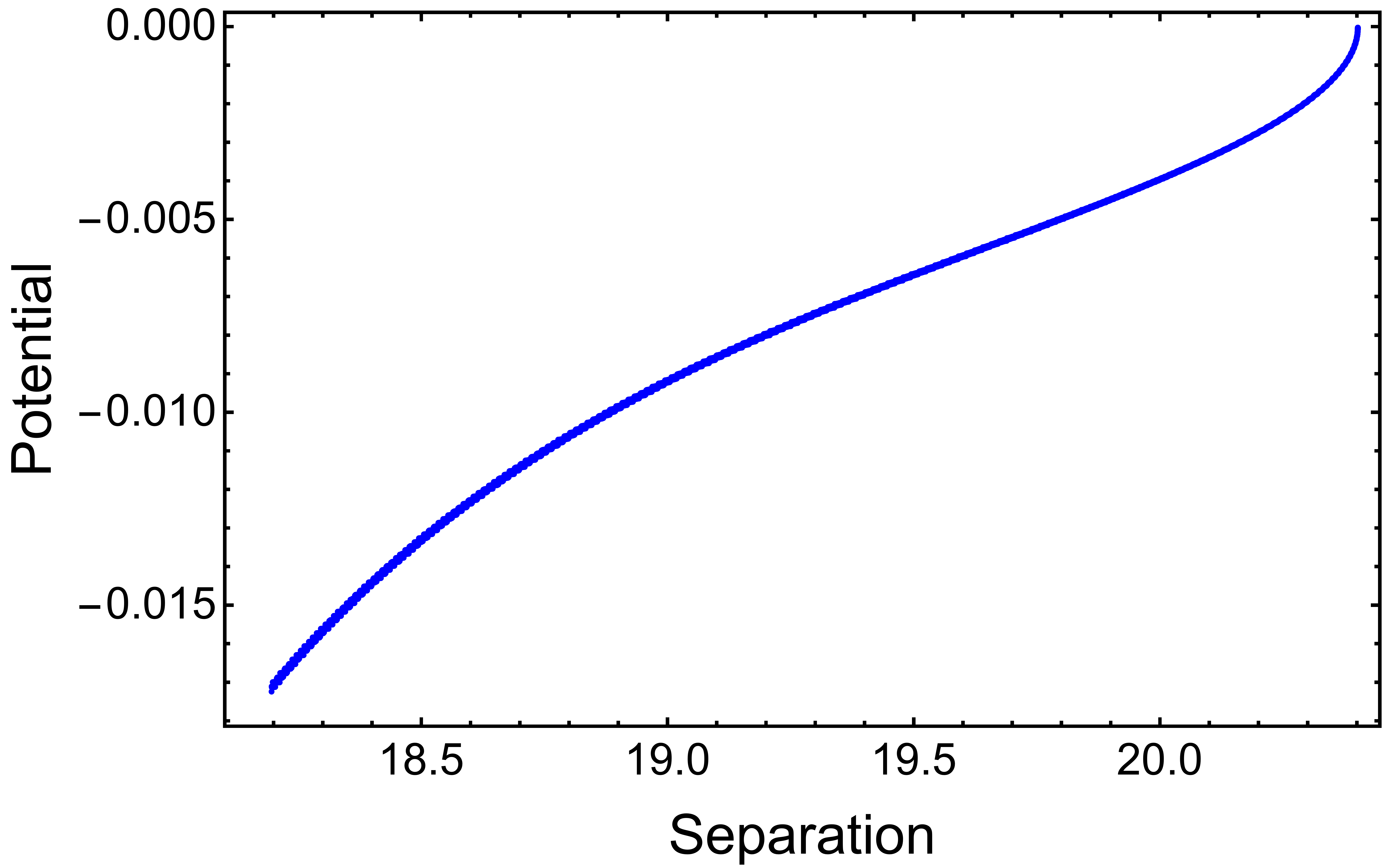}
\par\smallskip
{\small\textbf{(a)} In-phase interaction ($\chi=0$): attractive effective potential.}\\[0.5em]
\includegraphics[width=\columnwidth]{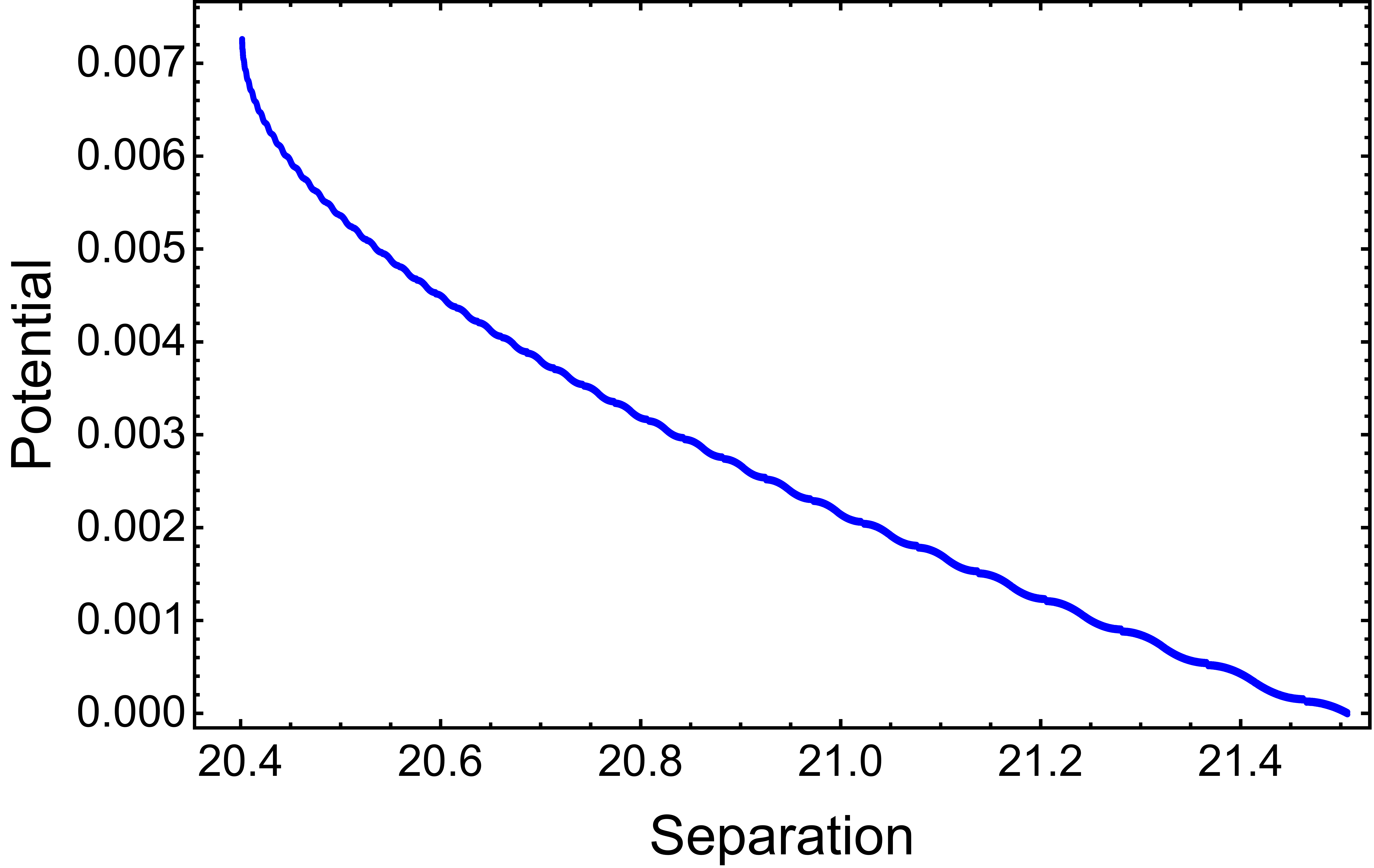}
\par\smallskip
{\small\textbf{(b)} Out-of-phase interaction ($\chi=\pi$): repulsive effective potential.}
\caption{Effective interaction potentials $V(s)$ extracted from real-time collision
dynamics of two FTSs. The potentials are shifted such that
$V(\infty)=0$.}
\label{fig:potentials}
\end{figure}

The resulting effective interaction potentials extracted from the collision
trajectories are shown in Fig.~\ref{fig:potentials}. In-phase collisions, $\chi=0$, where $\chi$ is the phase difference at collision, lead to an attractive effective potential, as shown in 
Fig.~\ref{fig:potentials}(a), while out-of-phase collisions, $\chi=\pi$, produce a repulsive effective potential,
Fig.~\ref{fig:potentials}(b). Consistent with Fig.~\ref{fig:potentials}, in-phase collisions lead to strong
overlap and, in many cases, soliton merger, whereas out-of-phase collisions
suppress overlap and favor elastic scattering.

In the strongly inelastic regime, the attractive effective interaction explains
\emph{how} the two solitons are driven into strong overlap, but it does not by
itself explain \emph{why} the resulting merged object remains stable instead of
rapidly fragmenting or dispersing. To address this point, we next examine the
energetic cost associated with the soliton interface. This leads naturally to an
interpretation based on interfacial energetics and a Young--Laplace--type
balance, whereby the bulk pressure inside the flat-top structure is balanced by
a finite interfacial line tension at its curved boundary. Such a balance provides
a physical mechanism for the stabilization of merged FTSs formed
after strongly inelastic collisions.

%

\section{Coalescence and long-lived merged states}
\label{sec:discussion_coalescence}

Some inelastic collisions lead to long-lived merged or permanently coalesced
states, in which the two FTSs do not re-separate after impact.
While a complete dynamical theory of permanent coalescence is beyond the scope
of the present collision-focused work, the persistence of merged states can be
understood quantitatively in terms of the interfacial (edge) energetics of
two-dimensional FTSs, together with a variational
characterization of their bulk structure. These elements provide a consistent
energetic framework for interpreting why strong overlap during inelastic
collisions relaxes toward a stable merged configuration rather than leading to
disintegration.


\subsection{Interfacial energetics and Young--Laplace balance}
\label{subsec:interfacial_energetics}

We summarize the energetic ingredients underlying the stabilization mechanism,
following the approach of Ref.~\cite{Novoa2009}. For stationary states
$\psi=u(r)e^{-i\mu t}$ at fixed chemical potential $\mu$, stability is governed by
the minimization of the grand potential,
\begin{equation}
\Omega[u] = E[u] - \mu N[u],
\label{eq:grand_potential}
\end{equation}
rather than by minimization of the energy $E$ alone. Substituting
Eqs.~\eqref{eq:norm} and \eqref{eq:total_energy} into
Eq.~\eqref{eq:grand_potential} yields
\begin{equation}
\begin{aligned}
\Omega[\psi]
&=
\iint
\Bigg[
\frac{1}{2}\left(|\psi_x|^{2}+|\psi_y|^{2}\right)
-\frac{g_{2}}{2}|\psi|^{4}
-\frac{g_{3}}{3}|\psi|^{6} \\&
-\mu |\psi|^{2}
\Bigg]
\,dx\,dy .
\end{aligned}
\label{eq:grand_potential_total_form}
\end{equation}

In two dimensions at zero temperature, the pressure is defined as the derivative
of the grand potential with respect to the transverse area,
$
p = -\left(\frac{\partial \Omega}{\partial S}\right)_{\mu},
$
where $S$ is the two-dimensional area of the system and the derivative is taken at
fixed chemical potential $\mu$. For stationary radially symmetric states, this
definition leads to the local pressure
\begin{equation}
p(r)
=
-\frac{1}{2}u_r^{2}
+\frac{g_{2}}{2}u^{4}
+\frac{g_{3}}{3}u^{6}
+\mu u^{2}.
\label{eq:pressure_radial}
\end{equation}

For FTSs, $p(r)$ is positive and nearly constant in the bulk
($p_c>0$), while it becomes negative within a narrow interfacial region where
gradient contributions dominate, cf.\
Fig.~\ref{fig:radialPressure_collisionPaper}. The corresponding line tension
(surface tension in three dimensions) is obtained as
$
\sigma = -\frac{1}{R}\int_R^\infty r\,p(r)\,dr,
$
where $R$ is the soliton radius. Minimization of $\Omega$ with respect to $R$
yields a two-dimensional Young--Laplace--type balance,
$
p_c \simeq \frac{\sigma}{R}.
$

\begin{figure}[t]
\centering
\includegraphics[width=\columnwidth]{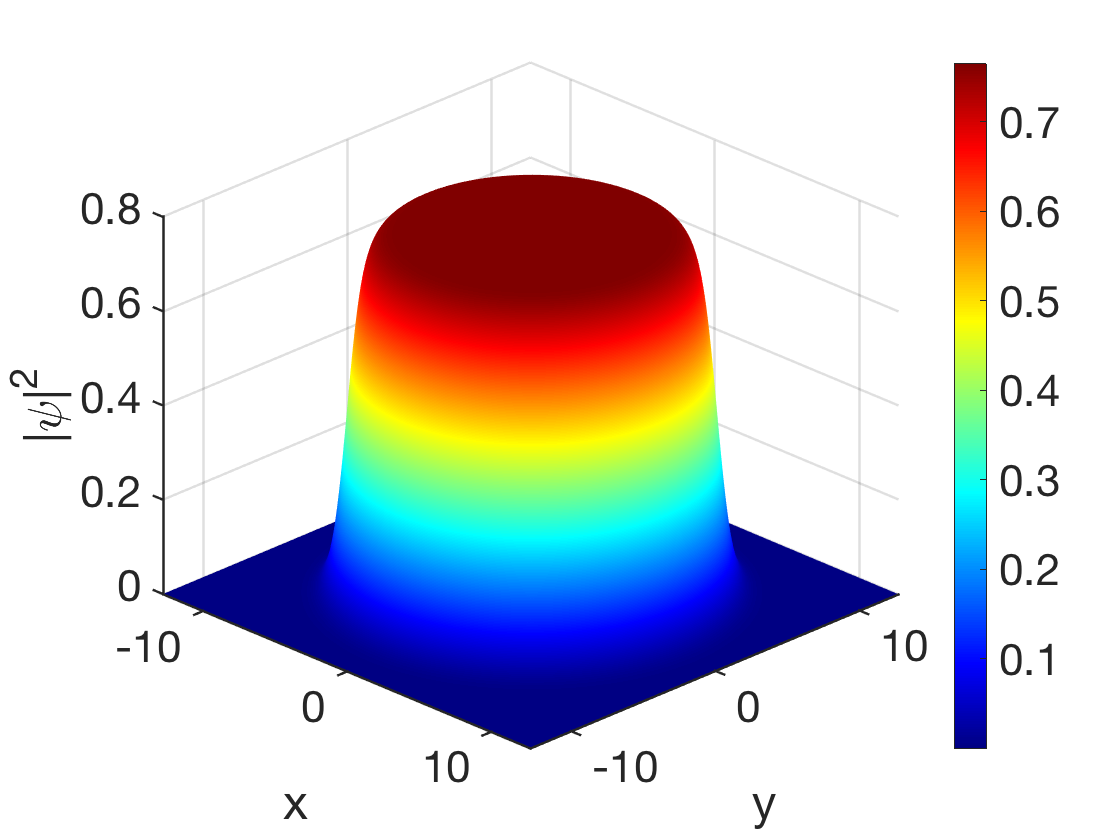}
\caption{Stationary single two-dimensional flat-top soliton obtained by ITP of Eq.~\eqref{eq:CQNLSE2D} with cubic--quintic nonlinearities
$g_2=4$ and $g_3=-4$, and norm $N=200$.
The computation converges to a ground state with chemical potential
$\mu\simeq-0.717$ and total energy $E\simeq-136.24$.
The soliton exhibits a nearly uniform bulk density with peak amplitude
$|\psi|_{\max}\simeq0.875$ and characteristic width
$\mathrm{FWHM}\simeq18.18$.
This stationary state serves as the reference configuration used to compute
the radial pressure profile and line tension shown in
Fig.~\ref{fig:radialPressure_collisionPaper}.}

\label{fig:singleFTS_reference}
\end{figure}

The stationary single flat-top soliton shown in
Fig.~\ref{fig:singleFTS_reference} is obtained from an independent
ITP run of Eq.~\eqref{eq:CQNLSE2D}, using the same
nonlinear parameters as in the collision simulations ($g_2=4$, $g_3=-4$),
but without a second soliton present. For this reference state, the ITP converges to a ground
state characterized by $N=200$, $E=-136.24$, and $\mu=-0.7171$. The
corresponding density profile exhibits a nearly uniform bulk interior with
a full width at half maximum $\mathrm{FWHM}\simeq18.18$ and a peak
amplitude $|\psi|_{\max}\simeq0.875$.

Using this stationary reference configuration, the radial pressure profile
$p(r)$ shown in Fig.~\ref{fig:radialPressure_collisionPaper} is computed
from the radially averaged density. From this profile, the bulk pressure
is found to be $p_c\simeq2.46\times10^{-2}$. The soliton radius, defined as
the location where the radial density drops most steeply, is
$R\simeq9.07$. The corresponding grand potential is $\Omega\simeq7.18$,
yielding a line tension $\sigma\simeq0.238$. These values satisfy the
Young--Laplace relation to good accuracy, with
$\sigma/R\simeq2.62\times10^{-2}$, consistent with the bulk pressure
extracted directly from $p(r)$.

\begin{figure}[t]
\centering
\includegraphics[width=\columnwidth]{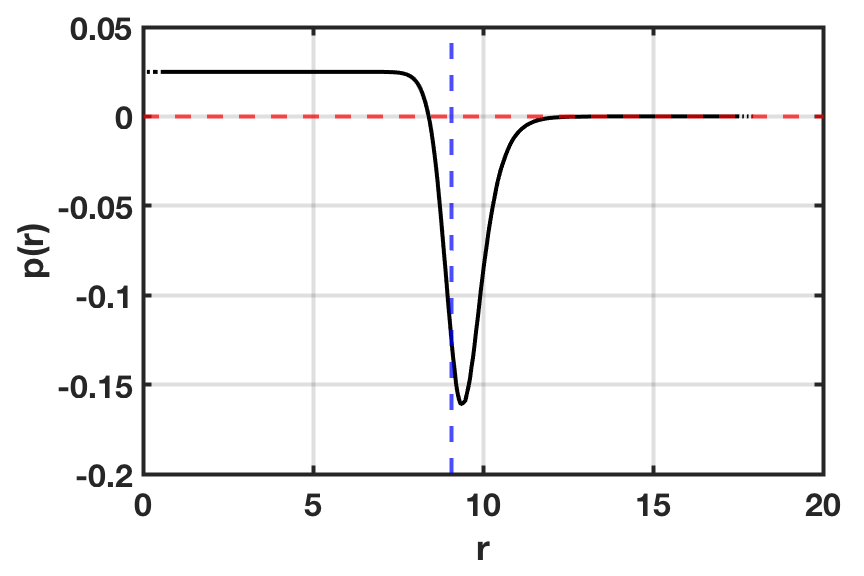}
\caption{Radial pressure profile $p(r)$ computed from the stationary single
flat-top soliton shown in Fig.~\ref{fig:singleFTS_reference}.
The pressure is positive and nearly constant in the bulk ($p_c$), becomes
negative in a narrow interfacial (edge) region dominated by gradients, and
approaches zero outside.
The dashed line marks the interface radius $R$, defined as the radius where
the density drops most steeply and used to compute the line tension $\sigma$.}
\label{fig:radialPressure_collisionPaper}
\end{figure}

%
\subsection{Variational energetic stability of merged states}

A second quantitative element is that the stationary FTS used in our simulations possess a robust energetic structure that can be captured by a compact variational baseline. Single-soliton ground states are radially symmetric and can therefore be parameterized by a super-Gaussian ansatz,
\begin{equation}
\psi(r) = A \exp\!\left[-\frac{1}{2}\left(\frac{r}{w}\right)^{2m}\right],
\end{equation}
with fixed norm $N$. Evaluating the conserved energy functional on this profile yields
\begin{multline}
E(w,m) =
\frac{g_{1} m^2 N}{\Gamma(1/m) w^2}
- \frac{g_2 m N^2}{2\pi\,2^{1/m}\Gamma(1/m) w^2} \\
- \frac{g_3 m^2 N^3}{3\pi^2\,3^{1/m}[\Gamma(1/m)]^2 w^4},
\label{eq:Ewm_discussion}
\end{multline}
where the amplitude follows from norm conservation,
\begin{equation}
A = \sqrt{\frac{N m}{\pi \Gamma(1/m)}}\,\frac{1}{w}.
\end{equation}


The variational parameters $(w,m)$ are determined by direct minimization of $E(w,m)$,
\begin{equation}
\frac{\partial E}{\partial w} = 0, \qquad
\frac{\partial E}{\partial m} = 0 .
\end{equation}
For representative parameters $g_{1}=1/2$, $g_2=4$, $g_3=-4$, and $N=100$, this procedure yields
\begin{equation}
w_0 = 6.643, \qquad
m_0 = 4.227, \qquad
A_0 = 0.8906 .
\end{equation}

\begin{figure}[t]
\centering
\includegraphics[width=0.9\linewidth]{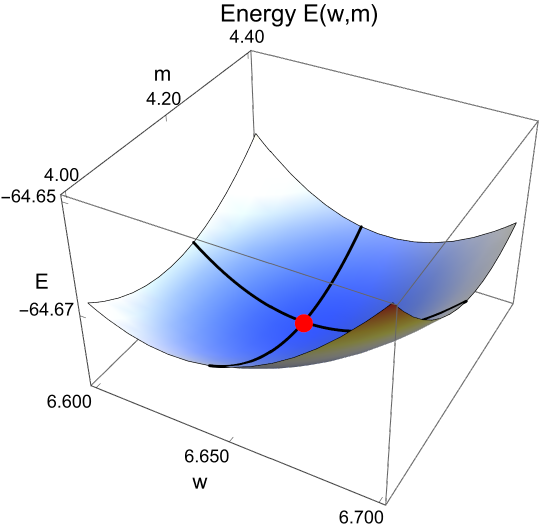}
\caption{
Energy surface $E(w,m)$ obtained by evaluating the variational energy functional
[Eq.~\eqref{eq:Ewm_discussion}] for the super-Gaussian ansatz at fixed norm.
The numerical parameters used are
$g_{1}=0.5$, $g_2=4$, $g_3=-4$, and $N=100$.
The marked minimum at $(w_0,m_0)=(6.643,\,4.227)$ corresponds to a variationally
stable flat-top soliton, as discussed in the text.
}
\label{fig:energy_landscape}
\end{figure}

The stationary point $(w_0,m_0)$ corresponds to a local minimum of the energy surface. This is confirmed by the positive definiteness of the Hessian matrix evaluated at this point,
\begin{equation}
\begin{aligned}
H &=
\begin{pmatrix}
\partial^2 E/\partial w^2 & \partial^2 E/\partial w\,\partial m \\
\partial^2 E/\partial w\,\partial m & \partial^2 E/\partial m^2
\end{pmatrix} \\
&=
\begin{pmatrix}
11.73 & -0.666 \\
-0.666 & 0.547
\end{pmatrix},
\end{aligned}
\end{equation}

with $\partial^2 E/\partial w^2 > 0$, $\partial^2 E/\partial m^2 > 0$, and $\det(H)=5.97>0$. The corresponding energy surface and its minimum are shown in Fig.~\ref{fig:energy_landscape}.

The existence of this local energetic minimum explains why, once strong overlap occurs during an inelastic collision, the system relaxes toward a stable merged configuration rather than fragmenting or dispersing. As a result, the two FTSs lose their individual soliton identities and coalesce into a single, long-lived structure. In this sense, coalescence represents a natural outcome of the combined bulk incompressibility and interfacial energetic cost inherent to two-dimensional FTSs.


We quantify the accuracy of the variational energy minimum by comparing the
variational energy evaluated at the optimal parameters $(w_0,m_0)$ with exact
stationary ground states obtained independently by ITP of Eq.~\eqref{eq:CQNLSE2D} at fixed norm.
For each value of the norm $N$, the ITP procedure converges to the minimum-energy
stationary solution, which serves as a numerical reference.

For a representative case with $N=100$, the exact ground-state energy obtained by
ITP is $E_{\mathrm{ITP}}=-65.1408$.
The energy-minimized variational ansatz (EVA), with optimal parameters
$w=6.643$ and $m=4.227$, yields $E_{\mathrm{EVA}}=-64.6915$, corresponding to a
relative deviation
$\epsilon = |E_{\mathrm{EVA}} - E_{\mathrm{ITP}}|/|E_{\mathrm{ITP}}|
\simeq 6.9\times10^{-3}$.
This level of agreement confirms that direct energy minimization provides an
accurate energetic description of stationary two-dimensional FTSs.

Table~\ref{tab:VA_norms} extends this comparison to a broader range of norms.
The persistence of small relative deviations across increasing $N$ demonstrates
that the variational minimum represents a robust energetic feature rather than a
fine-tuned solution. This robustness supports the interpretation that, once
strong overlap is induced during an inelastic collision, the system relaxes
toward a stable merged configuration associated with this energetic basin rather
than fragmenting or dispersing.

\begin{table}[t]
\centering
\caption{
Comparison of variational and exact stationary energies for different norms $N$.
For each $N$, $E_{\mathrm{ITP}}$ is obtained from ITP, while
$E_{\mathrm{VA}}$ is evaluated at the corresponding variational energy minimum. The relative deviation $\epsilon$ is defined in the text.}
\label{tab:VA_norms}
\begin{tabular}{cccc}
\toprule
$N$ & $E_{\mathrm{ITP}}$ & $E_{\mathrm{VA}}$ & $\epsilon$ \\
\midrule
100 & $-65.1408$  & $-64.6915$  & $6.90\times10^{-3}$ \\
140 & $-93.4132$  & $-92.8441$  & $6.09\times10^{-3}$ \\
160 & $-107.6437$ & $-107.0200$ & $5.80\times10^{-3}$ \\
180 & $-121.9216$ & $-121.2470$ & $5.53\times10^{-3}$ \\
200 & $-136.2398$ & $-135.5150$ & $5.32\times10^{-3}$ \\
\bottomrule
\end{tabular}
\end{table}

To further support this energetic interpretation, we performed a direct numerical comparison between the energy of the coalesced flat-top soliton formed after an inelastic collision and the combined energy of the two initially separated solitons before collision. Starting from a stationary ITP configuration of two well-separated flat-top solitons with equal norms $N_L = N_R = 100$, we obtain a total pre-collision energy $E_{\mathrm{two}} = E_L + E_R = -130.26$. After a strongly inelastic collision, we calculate the energy of the bound post-collision state, excluding the emitted radiation. This coalesced core has a slightly reduced norm $N_{\mathrm{core}} =196.09$, due to radiative losses during the collision, and a lower energy $E_{\mathrm{core}} = -130.88$. Crucially, the inequality $E_{\mathrm{core}} < E_{\mathrm{two}}$ holds, with a difference $\Delta E = -0.62$, demonstrating that the merged configuration is energetically favored over two separated flat-top solitons. This explains why the solitons remain coalesced after collision. In contrast to the integrable one-dimensional cubic NLSE, where the energy of two solitons is simply the sum of their individual energies and a merged state can split back into two solitons without any energetic penalty, the cubic--quintic model considered here is nonintegrable, and radiation losses allow the system to relax into a lower-energy bound state. This numerical energy comparison therefore confirms that coalescence in two-dimensional flat-top solitons is energetically driven rather than a transient dynamical effect.


%

\section{Conclusions}
\label{sec:conclusions}

We have investigated elastic, inelastic, and coalescent collisions of
two-dimensional FTSs supported by the cubic--quintic nonlinear
Schr\"odinger equation. Several main conclusions emerge from this study.

First, we find that despite the intrinsic nonintegrability of the two-dimensional
system, flat-top soliton collisions can exhibit well-defined regimes of nearly
elastic and inelastic scattering. As in earlier studies of one-dimensional
flat-top soliton collisions, the transition between elastic and inelastic
regimes is primarily controlled by the relative phase of the solitons at the
collision point. Out-of-phase collisions suppress overlap and radiation
emission, leading to nearly elastic scattering, whereas in-phase collisions
promote strong overlap and energy redistribution, resulting in inelastic
behavior. Kinetic-energy diagnostics provide a clear and quantitative
characterization of this phase-controlled transition.

Second, we have shown that the dependence of collision outcomes on initial
separation and imposed initial phase arises from deterministic phase accumulation
during propagation. By varying the initial separation, one effectively tunes
the relative phase at impact, leading to alternating windows of elastic and
inelastic scattering. This phase-controlled structure persists in two
dimensions despite the presence of transverse degrees of freedom and
additional radiation channels, underscoring the central role of phase
interference in organizing flat-top soliton collision dynamics.

Third, a subset of strongly inelastic collisions leads to long-lived merged or
permanently coalesced states. We have demonstrated that the persistence of
these merged structures reflects the energetic properties of two-dimensional
FTSs rather than a transient dynamical effect. In particular, direct energy
comparisons show that the bound post-collision core has a lower energy than the
sum of the energies of the two separated flat-top solitons before collision,
confirming that coalescence is energetically favored. Effective interaction
potentials extracted from collision trajectories show that in-phase solitons
experience a net attractive interaction that drives strong overlap. The
subsequent stability of the merged object is explained by interfacial
energetics: a Young--Laplace--type balance between bulk pressure and interfacial
line tension provides a natural mechanism for stabilizing the post-collision
state. A variational analysis based on direct energy minimization
supports this picture. The existence of a robust local minimum in the
variational energy landscape explains why, once strong overlap occurs, the
system relaxes toward a stable merged configuration rather than fragmenting or
dispersing. In this sense, coalescence emerges as a consequence of the combined
bulk incompressibility and interfacial energetic cost inherent to
two-dimensional FTSs.

Finally, the results reported here highlight the importance of interfacial
energetics in higher-dimensional flat-top soliton dynamics. The combined use
of collision diagnostics, effective interaction forces, and energetic
characterization provides a coherent framework for understanding elastic,
inelastic, and coalescent scattering processes. This framework can be extended to explore the role of surface tension in more
complex geometries, including anisotropic FTSs, externally confined systems,
non-head-on collision scenarios, and interactions among multiple solitons in two
dimensions.

\begin{acknowledgments}
M.~O.~D.~Alotaibi acknowledges support from the Kuwait Foundation for the Advancement of Sciences (KFAS) under the KFAS--ICTP Kuwait Programme and the Kuwait Visiting Scientists Scheme Fellowship.
L.~Al~Sakkaf acknowledges support from the ICTP--Arab Fund Associates Programme (ARF01--AFESD Grant No.~14/2023) under the project ``Advancing the Capabilities of Arab Researchers and Students''.
\end{acknowledgments}


\end{document}